\documentstyle[12pt]{article}
\begin{document}
\begin{titlepage}
\pagestyle{empty}
\baselineskip=21pt
\rightline{Alberta-THY-18/96}
\rightline{hep-ph/9606463}
\rightline{June 1996}
\vskip .2in
\begin{center}
{\large{\bf  Cosmological Reheating and Self-Interacting Final State  
Bosons}} \end{center}
\vskip .1in
\begin{center}
Rouzbeh Allahverdi
  and
Bruce A. Campbell

{\it Department of Physics, University of Alberta}

{\it Edmonton, Alberta, Canada T6G 2J1}

\vskip .1in

\end{center}
\vskip .5in
\centerline{ {\bf Abstract} }
\baselineskip=18pt
We consider inflaton decay to final state
bosons with self-interactions of moderate strength.  We
find that such final state self-interactions
qualitatively alter the reheat dynamics.
In the case of  narrow-band resonance decay, where a quantitative  
analysis is possible, we show that these final state interactions  
regulate the decay rate.
The phenomenon of parametric amplification is then effectively  
suppressed, and does not drastically enhance the decay rate and  
reheat temperature.  Detailed applications of our results to  
realistic classes of inflationary models  will be considered  
elsewhere.
\noindent
\end{titlepage}
\baselineskip=18pt

\def\la{~\mbox{\raisebox{-.7ex}{$\stackrel{<}{\sim}$}}~}
\def\ga{~\mbox{\raisebox{-.7ex}{$\stackrel{>}{\sim}$}}~}1
\def\mtw#1{m_{\tilde #1}}
\def\tw#1{${\tilde #1}$}
\def\beq{\begin{equation}}
\def\eeq{\end{equation}}

Inflationary cosmology  \cite{1}, provides solutions
to the flatness, isotropy and stable relic
problems of  the standard hot big bang. In these models the universe  
experiences a period of
superluminal expansion during which its energy density is dominated
by the potential energy of a scalar field (inflaton).  Inflation ends
when the inflaton enters the oscillatory regime
during which we have a matter-dominated FRW-universe, after which
the inflaton decays to relativistic particles (reheating).  Reheating  
represents the crucial transition from the epoch  of scalar-field  
dominated dynamics to a hot FRW universe.
After reheating, the universe becomes radiation-dominated
and its evolution is just that of the standard hot big bang .

In the standard picture of reheating \cite{1,2} the effective decay  
of the inflaton
occurs when $\Gamma \simeq H$, where $\Gamma$ is the
one particle decay rate, and the reheat temperature of the universe  
is ${T}_{R} \sim 0.1{(\Gamma)}^{{1\over 2}}$ (from now on  
${M}_{pl}=1$ and all dimensionful quantities are expressed in these  
units unless otherwise indicated).  In this picture, which we will  
refer to as  the linear regime, perturbation  theory is taken to be  
valid
and the occupation number of the final state bosons is assumed to be  
smaller than one (for fermions
this is assured because of Pauli blocking).  It has recently been  
recognized
within different approaches \cite{3} -\cite{31}
that this picture is incomplete, and nonlinear effects can change it  
essentially,
leading by parametric amplification to an enhanced decay of the  
inflaton (for a recent review see, e.g. \cite{3}).  There are two  
different possible regimes of such a parametrically amplified decay.  
In the first the decay occurs over many oscillation times of the  
inflaton, but experiences parametric amplification with moderately  
large occupation numbers for the modes of the decay product field;  
this "narrow-band resonance" case is amenable to analytic  
calculation, and we will be able to analyze the modifications to  
parametric amplification due to final state-self interactions of the  
decay products quantitatively. In the second case of "broad-band  
resonance" the amplification of the decay is so strong that there is  
explosive decay of the inflaton field on a time-scale not   
hierarchically longer than the oscillation time, with large  
occupation numbers for the modes of the decay product field; this  
scenario is more difficult to analyze quantitatively, but the  
physical mechanisms which we discuss are of sufficient generality  
that we expect that they alter the decay dynamics in this case also.

The consequences
of a parametrically amplified decay could be significant, and pose  
major difficulties in the construction of viable inflationary models.   
If the decay is
very fast, the final state particle modes have, in general, very  
large occupation numbers and
are far from thermal equilibrium; these  fluctuations have certain  
effects similar to
very high temperature  thermal corrections \cite{9,12} , and after  
thermalization may themselves lead to high reheat temperatures, with  
thermal energy densities of order the inflationary energy density.   
While on the one hand GUT symmetry restoration  due to these  
non-thermal configurations
revives   GUT scenarios for  baryogenesis \cite{23,16,17} , on the  
other hand it reintroduces the problem of heavy topological defects  
whose solution was one of the initial motivations for inflation.

Furthermore, realistic models seeking to implement inflation need to  
be able to stabilize the required flat potential against radiative  
corrections.  The only presently known method to achieve this in the  
presence of gauge,  scalar potential, Yukawa, and gravitational  
interactions is by using (approximate) supersymmetry to enforce the  
appropriate non-renormalization. Parametrically amplified decay of  
the inflaton, and the resulting efficient reheat, poses a mortal  
danger to realistic, supersymmetric, inflationary models.  In  
supersymmetric theories
 the reheat temperature  is constrained  \cite{32,33,34} by the need  
to avoid thermal overproduction of gravitinos. This leads to an upper  
bound on the reheat temperature of order $10^{8} - 10^{9}  GeV$,  
which is orders of magnitude below what would be achieved by  
efficient reheat from amplified inflaton decay.  One obvious way to  
avoid this disaster would be to insure that the inflaton is   
sufficiently weakly coupled to its bosonic decay products that the  
decay never experiences parametric amplification. This is typically  
the case for inflatons in a  hidden sector which have only  
gravitational strength couplings to their (observable sector) decay  
products. On the other hand, in many models one introduces direct  
superpotential couplings of the inflaton to the chiral scalars into  
which it decays, and  we wish to consider the nature of the reheat  
dynamics in such models, and whether they are ruled out by  
constraints on reheating.

In particular, we will examine the effects that the final state  
self-interactions of the decay products have on the parametric  
amplification of the inflaton decay. Because of the large occupation  
numbers for the modes of the produced decay bosons, we expect that  
the presence of self-interactions of these bosons will result in  
large effective masses being induced for these modes.  If the bosons  
are thermalized these may be interpreted as thermal plasma masses  
from self-interaction; more generally they will occur as induced  
effective plasma mass terms in the mode equations for the decay  
field. As the mode occupation numbers increase, so do these induced  
masses,  until they equal the mass of the inflaton, cutting off the  
decay. Decay resumes with the thermalization of the decay products,  
and their dilution and redshift by cosmic expansion, such that their  
induced masses dip below the inflaton mass; the system thus proceeds  
in a quasi-stationary process of decay and dilution such that the  
induced mass of the decay products is always of order the inflaton  
mass. This regulates the parametric amplification of the decay,  
preventing abrupt and efficient reheat.  In our discussion we will   
analyze the effects of final state self-interactions for inflaton  
decays that would otherwise be in the regime of narrow-band  
parametric amplification.  However, because the dominant effect is a  
kinematical cutoff of the decay, due to the self-induced plasma mass  
of the final state decay products arising from their strong  
self-interaction, we expect similar effects in the case of broad-band  
resonance. Indeed, recent studies of the broad-band decay regime  
\cite{7,10,14} indicate that the explosive  decay  from the sequence  
of higher resonance bands can only effectively produce particles  
whose mass does not exceed that of the inflaton by more than an order  
of magnitude.  So in this case too there will be a kinematical cutoff  
due to the final-state self-interaction induced plasma mass of the  
decay products; although it will now be regulated to be not more than  
of order ten times the inflaton mass the qualitative effects should  
be otherwise similar.
Applications of our present analysis  to realistic classes of  
supersymmetric inflationary models will be considered  elsewhere   
\cite{31}.

As a basis for our subsequent arguments we will consider a chaotic  
inflation model with the following potential, whose features we take  
to resemble the generic features of scalar potentials which arise in  
supersymmetric theories:
\beq
V={1 \over 2}{m}^{2}{\phi}^{2} + \sigma\phi {\chi}^{2}  + {h}^{2}  
{\phi}^{2}{\chi}^{2} +{g}^{2}{\chi}^{4}
\eeq
where for schematic simplicity the inflaton $\phi$ and the matter  
scalar $\chi$ are taken to be real scalar fields and the  
self-coupling of the $\chi$ field is considered to be of moderate  
strength
${10}^{-1}< {g}^{2} < 1$.  In supersymmetric theories where the decay  
scalars are standard model chiral scalars which are gauge  
non-singlets the quartic potential terms in $ \chi$ arise as D-terms  
and the coupling is of gauge coupling strength  $g^{2}$. For  
inflaton-scalar couplings arising from superpotentials in  
supersymmetric theories one has $ \sigma = 2 h m $, and the cubic and  
quartic couplings  of the inflaton to $\chi$ are related to each  
other. In general the superpotential couplings $ h$ may be, and in  
viable supersymmetric inflationary models are usually chosen to be,  
much smaller than gauge couplings $ h \ll g$. The inflaton mass $m$  
must be bounded  by  $m{\  
\lower-1.2pt\vbox{\hbox{\rlap{$<$}\lower5pt\vbox{\hbox{$\sim$}}}}\ }  
{10}^{-6}$  in order to be consistent with COBE data on microwave   
background fluctuations \cite{35}.

First let us review the effects of parametric amplification on  
inflaton decay, ignoring self-interaction of the decay products. The  
nonlinear effects that lead to
amplified decay act in two different regimes: \\
$[1]~~$   ${{m}^{4} \over {\sigma}^{2}}{\  
\lower-1.2pt\vbox{\hbox{\rlap{$<$}\lower5pt\vbox{\hbox{$\sim$}}}}\ }  
\phi {\  
\lower-1.2pt\vbox{\hbox{\rlap{$<$}\lower5pt\vbox{\hbox{$\sim$}}}}\ }  
{{m}^{2} \over \sigma}$ for the cubic coupling
and ${{m}^{4} \over {h}^{4}}{\  
\lower-1.2pt\vbox{\hbox{\rlap{$<$}\lower5pt\vbox{\hbox{$\sim$}}}}\  
}{\phi}^{3}{\  
\lower-1.2pt\vbox{\hbox{\rlap{$<$}\lower5pt\vbox{\hbox{$\sim$}}}}\  
}{{m}^{3} \over {h}^{3}}$ for the quartic coupling.  This is the  
narrow-band resonance case which can be analyzed perturbatively, and  
the dominant effect is the
large occupation number for $\chi$'s.  This case has been considered  
in
\cite{7,8,10,11,18} , where it is shown that parametric amplification  
occurs and there are narrow-band resonances for $\chi$ production
at $k={m \over 2}$ and $k=m$ for the cubic and quartic couplings  
respectively.  This is the case for which we will  make quantitative  
estimates of the effect of inclusion of final state self-interaction  
of the decay products. \\
$[2] ~~$  $\phi {\  
\lower-1.2pt\vbox{\hbox{\rlap{$>$}\lower5pt\vbox{\hbox{$\sim$}}}}\  
}{{m}^{2} \over \sigma}$ for the cubic coupling and $\phi{\  
\lower-1.2pt\vbox{\hbox{\rlap{$>$}\lower5pt\vbox{\hbox{$\sim$}}}}\  
}{m\over h}$ for the quartic coupling.  Here perturbation theory
is not valid and problem is highly nonlinear.  In this case there is
broad-band resonance for a large domain of momenta \cite{7,10} that  
leads to
an explosive decay of the inflaton.  As noted above, because of the  
essentially kinematic nature of the cutoff we expect the final-state  
self-interaction effects in this case to be qualitatively similar to  
those in the narrow band case.

Note that if the cubic and quartic couplings of the inflaton to the  
decay scalar are of the form arising from a superpotential coupling  
[$ \sigma\simeq 2 h m $], then the conditions for parametric  
amplification are more general for the cubic coupling and it will  
dominate the decay. Another important observation is that although  
most inflatons may decay during a stage of amplified decay this does  
not lead to the decay of the entire energy density of the inflaton.   
In the case of narrow-band resonance the decay stops no later than  
the time when $\phi$ has the minimum value in the abovementioned  
range (${{m}^{4} \over {\sigma}^{2}}$ for the cubic coupling and  
${({m \over h})}^{{4 \over 3}}$ for the quartic coupling) \cite{7,8},  
while in the case of broad-band resonance it stops at the time of the  
transition
from broad-band to narrow-band resonance
when the decay becomes out of equilibrium \cite{7,10,25}.  After
that, the remaining energy density
of the inflaton is redshifted as ${R}^{-3}$ where energy density of
relativistic $\chi$'s is redshifted as ${R}^{-4}$.  The decay of the  
inflaton will then be
completed as in the usual  picture, and if the energy density of the  
inflaton dominates at that time
there will be significant dilution of relic densities from the first  
stage of reheat \cite{25,16}.  In supersymmetric theories this  
dilution is not, generally, sufficient by itself to solve the problem  
of overproduction of gravitinos.  However, combined with the effect  
of final state self-interaction, which we consider below, it can  
successfully resolve the gravitino problem in many models.

Now let us consider generally the changes to the parametric  
amplification reheat dynamics that arise from the self-interactions  
of the final state decay products.   Since by assumption the inflaton  
decays to observable sector standard model (s)particles the final  
state bosons carry gauge quantum numbers.
For these fields self-interactions with couplings
as strong as the gauge coupling arise
at tree-level from D-terms in supersymmetric models
\footnote{There are, of course, directions in scalar field space  
which are both D-flat and F-flat in the supersymmetric standard  
model; in realistic no-scale supergravity models they will in general  
already have developed Planck scale vevs during inflation \cite{36}.  
If the decay coupling of the inflaton happened to align along one of  
these directions then final state interactions of the type we  
consider would not affect this particular decay mode. More generally  
the decay products will themselves, in turn, have decay modes along  
these directions. We consider these issues elsewhere \cite{31}.}, 

and at the one-loop level in the
non-supersymmetric case.  In addition,
there are couplings of non-singlet scalars to the gauge fields as  
well as possible large superpotential couplings.  The decay of the  
inflaton
produces quantum fluctuations along
the direction of final state particles in field space and drives them  
to large field values where the
effect of self-coupling becomes important.  These large field values
induce a self-mass for the decay products
that subsequently slows down the decay and tends to shut it off .

  During the oscillatory regime
$\phi \cong{\phi}_{0} \cos{(mt)}$, and
immediately after the end of inflation ${\phi}_{0}\sim 10^{-1}$.  
Subsequently  ${\phi}_{0}$  decreases
with time because of
decay and Hubble expansion.  For ${\phi}_{0}{\  
\lower-1.2pt\vbox{\hbox{\rlap{$<$}\lower5pt\vbox{\hbox{$\sim$}}}}\ }  
{m \over h}$ the inflaton decays
predominantly via the cubic term.
As was discussed in \cite{7,8,10}, in the range
${({m \over h})}^{2}{\  
\lower-1.2pt\vbox{\hbox{\rlap{$<$}\lower5pt\vbox{\hbox{$\sim$}}}}\ }  
{\phi}_{0}{\  
\lower-1.2pt\vbox{\hbox{\rlap{$<$}\lower5pt\vbox{\hbox{$\sim$}}}}\ }  
{m \over h}$
parametric amplification occurs and there is a narrow-band resonance
for $\chi$ production at
$k={m \over 2}$.  In this range for the inflaton field we can
use the Feynman diagrams for one-particle decay, but  the occupation  
number
of final state particles is
non-trivial and must be taken into account in the calcualtions.   
According to \cite{7,8,10}
parametric amplification will
effectively convert most of the energy density of the inflaton into  
radiation once it is in the
afore-mentioned range.
Assuming rapid subsequent thermalization this leads to a reheat  
temperature
${T}_{R}\sim m{h}^{-{1 \over 2}}$
that is much higher than that of the usual picture $\sim 0.1{m}^{{1  
\over 2}}h$ for
reasonable values of $h$.  The
quartic self-coupling of $\chi$ will induce a finite temperature  
correction to the mass-squared of $\chi$, if they are thermalized, of  
order ${g}^{2}{{T}_{R}}^{2}\sim {g}^{2}{m}^{2}{h}^{-1}$ at this time
which is much larger than ${m}^{2}$ (as we mentioned above the  
non-thermal corrections that exist before thermalization are even  
larger).  However at an earlier time ${t}_{d}$ when the thermal (or  
non-thermal
\footnote{Whether the thermal or non-thermal correction should be  
considered depends on how rapid the thermalization rate is.  Details  
will be discussed shortly.}
) correction is  of order ${{m}^{2} \over 4}$
the one-particle decay becomes kinematically forbidden (note that the  
thermal correction to mass-squared of $\phi$ is of order ${h}^{2}  
{{T}_{R}}^{2}$ which is normally smaller than ${m}^{2}$ for $h$ a  
typical Yukawa type coupling).

The Hubble expansion
that subsequently occurs redshifts the correction to the mass-squared  
of $\chi$ as ${R}^{-2}$
and this causes further decay.  As long as ${t}_{d}<{H}^{-1}$ these  
successive steps of
expansion, decay and (perhaps) thermalization
continue.  Eventually the decay
is not effective enough to compensate for expansion and there is a  
delay before  the remaining energy
density of the inflaton is converted into relativistic particles as  
in the usual
picture, and the decay of the inflaton is completed.

Now let us perform a detailed mode by mode analysis of the effect of  
quartic self-coupling of
final state particles.  Consider the potential

\beq
\mbox{V}={1 \over 2}{m}^{2}{\phi}^{2}+2hm\phi {\chi}^{2}
\eeq
where $\phi$ and $\chi$ are both real scalars.  By mode expansion of
$\chi$ we derive the following equation for each mode

\beq
\ddot{{\chi}_{k}}+3{{\dot{R}} \over  
{R}}\dot{{\chi}_{k}}+({k}^{2}+4hm{\phi}_{0}\cos{mt}){\chi}_{k}=0
\eeq
where a dot denotes differentiation with respect to time, and in this  
equation for the modes associated with comoving wavenumbers, we are  
using the physical wavenumber $k$, where $R k = k_{comoving}$ with  
$R$ the scale factor .  For the moment we will ignore
the effect of Hubble expansion, since it generally occurs over a  
longer time scale than that of the effects we consider.   We will  
consider issues of thermalization and Hubble expansion below [for  
other treatments of the effect of Hubble expansion on the parametric  
resonance see \cite{7,8,11}]. By choosing $z={m \over 2}t$, and in  
the absence of final state self-interactions we derive a Mathieu  
equation for the modes of the $ \chi$ field
\beq
{{\chi}_{k}}^{"}+({{k}^{2} \over {({m \over 2})}^{2}} +  
{4hm{\phi}_{0} \over {({m \over 2})}^{2}}\cos{2z}){\chi}_{k}=0
\eeq
where prime denotes differentiation with respect to $z$.  In the case  
of narrow-band resonance (in which we perform our calculations) the  
Mathieu equation has resonance solutions in the first instability  
band ${({m \over 2})}^{2}-4hm{\phi}_{0}<{k}^{2}<{({m \over  
2})}^{2}+4hm{\phi}_{0}$.  Modes in this band grow exponentially in  
time, which one interprets as particle production.  The (slow) Hubble  
expansion will eventually drive the modes out of the instability  
band, but they spend enough time there to reach a substantial  
occupation number \cite{18}.  This time ${t}_{b}$ can be  
approximately calculated from $\delta k \sim k H {t}_{b}$ where $k={m  
\over 2}$ and $\delta k=8h{\phi}_{0}$ is the width of the first  
instability band which gives ${t}_{b} \sim {16 h \over {m}^{2}}$.

The quartic self-coupling ${g}^{2}{\chi}^{4}$ will induce an  
effective mass-squared ${{m}^{2}}_{eff}$ for the $\chi$ field and the  
Mathieu equation with the addition of this self-coupling will be  
modified to become

\beq
{{\chi}_{k}}^{"}+({{k}^{2} + {{m}^{2}}_{eff} \over {({m \over  
2})}^{2}} + {4hm{\phi}_{0} \over {({m \over  
2})}^{2}}\cos{2z}){\chi}_{k}=0
\eeq

 In a background of isotropically distributed $\chi$'s over a narrow  
band of momenta (which is the case in $\phi \rightarrow \chi \chi$  
decay before thermal distribution is achieved) ${{m}^{2}}_{eff} \sim  
{g}^{2} {{n}_{\chi} \over {E}_{\chi}}$ to the leading order,
with ${n}_{\chi}$ the number density of $\chi$'s and ${E}_{\chi}$  
their energy \cite{7,10,12,13,14}.  After thermalization, in a  
thermal background of temperature $T$,  we will have the standard  
result ${{m}^{2}}_{eff} \sim {g}^{2} {T}^{2}$.

At $t=0$ when $\phi$ starts oscillating ${{m}^{2}}_{eff}=0$ and
there is resonance in the band ${({m \over  
2})}^{2}-4hm{\phi}_{0}<{k}^{2}<{({m \over 2})}^{2}+4hm{\phi}_{0}$.   
With particle production in this band ${{m}^{2}}_{eff}$ increases and  
resonance in this band stops
when ${{m}^{2}}_{eff} = 8hm{\phi}_{0}$ after which the number density  
of particles produced in this band remains unchanged.  There will be,  
however, resonance in the band
${({m \over 2})}^{2}-12hm{\phi}_{0}<{k}^{2}<{({m \over  
2})}^{2}-4hm{\phi}_{0}$
that also stops when ${{m}^{2}}_{eff}$ increases by a further amount  
of
$8hm{\phi}_{0}$.  This
incremental change of ${{m}^{2}}_{eff}$ and smooth transition from  
one resonance band
to the next one continues until ${{m}^{2}}_{eff}={({m \over  
2})}^{2}$,
after that there is no resonance solution for physical states, i.e.
states with ${k}^{2}>0$.  If we take ${E}_{\chi}\simeq ({m \over 2})$  
for
particles in all bands a change of $8hm{\phi}_{0}$ in  
${{m}^{2}}_{eff}$ correponds to
an increase in number density of $\chi$'s by ${\delta n}_{\chi}\sim  
{4h{m}^{2}{\phi}_{0} \over {g}^{2} }$.  For the first band ${({m  
\over 2})}^{2}-4hm{\phi}_{0}<{k}^{2}<{({m \over  
2})}^{2}+4hm{\phi}_{0}$ and for $h{\phi}_{0}\ll m$ the occupation  
number ${f}_{{m \over 2}}$ required to shift the resonance to the  
next band is calculated to be

\beq
{\delta n}_{\chi}\sim {1 \over {(2\pi)}^{3}}{f}_{{m \over 2}}\times  
4\pi {({m \over 2})}^{2}\times 8h{\phi}_{0}\sim {4 h{\phi}_{0}{m}^{2}  
\over {g}^{2}}\Rightarrow {f}_{{m \over 2}}\sim {4{\pi}^{2} \over  
{g}^{2}}
\eeq

It has been shown \cite{18} that in the small amplitude limit the  
analytic result of the decay rate in the $n$-th instability band can  
also be derived from a physical process, $n$ particles with zero  
momentum that comprise the classical homogeneous inflaton field  
annihilating  into two final state bosons.  In particular, for the  
first instability band and for ${f}_{k}\ll 1$ we can use the  
one-particle decay rate from the standard perturbation theory $\Gamma  
={{h}^{2}m \over 8\pi}$.  Narrow-band parametric amplification has  
been shown to be indeed an induced process \cite{18} which means for  
${f}_{k}{\  
\lower-1.2pt\vbox{\hbox{\rlap{$>$}\lower5pt\vbox{\hbox{$\sim$}}}}\ }  
1$ the enhancement of decay rate by the factor $(1+ {f}_{k})$ must be  
taken into account in the particle calculation.  The time ${t}_{1}$  
which is needed to reach ${f}_{{m \over 2}} = 1$ can be approximately  
calculated as
\footnote{Note that ${e}^{-\Gamma {t}_{1}}\simeq 1-\Gamma {t}_{1}$  
for $\Gamma {t}_{1}\ll 1$, which is valid here.}

\beq
{{h}^{2}m \over 8\pi}{t}_{1}{m}{{\phi}_{0}}^{2}\sim  
{{2h{m}^{2}{\phi}_{0} \over {g}^{2}} \over {4{\pi}^{2} \over  
{g}^{2}}} \Rightarrow {t}_{1} \sim {4 \over \pi h {\phi}_{0}}
\eeq
and considering the enhancement factor the time $\delta t$ which is  
needed to reach ${f}_{{m \over 2}}={4{\pi}^{2} \over {g}^{2}}$ is

\beq
\delta t\sim {4 \over \pi h {\phi}_{0}} \ln{{4{\pi}^{2} \over  
{g}^{2}}}
\eeq

This is the time for the resonance in the band ${({m \over  
2})}^{2}-4hm{\phi}_{0}<{k}^{2}<{({m \over 2})}^{2}+4hm{\phi}_{0}$ to  
stop.  For the next bands ${k}^{2}$ is smaller and this means that  
less phase space volume is available for decay products or,
equivalently, the occupation number for those bands is larger.   
Therefore, more time will be needed for production of ${\delta  
n}_{\chi}\sim {4h{m}^{2}{\phi}_{0} \over {g}^{2} }$ in bands with  
smaller ${k}^{2}$, but not greatly so, as the larger occupation  
numbers are obtained rapidly due the the large coherent final state  
enhancement.  A reasonable lower estimate for the decay time  
${t}_{d}$ to effectively achieve ${{m}^{2}}_{eff} \sim {{m}^{2} \over  
4}$ is

\beq
{t}_{d}\sim {{{m}^{2} \over 4} \over 8 h m {\phi}_{0}} \delta t \sim  
{ m\over 8 \pi {h}^{2}{{\phi}_{0}}^{2}}\ln{{4 {\pi}^{2} \over  
{g}^{2}}}
\eeq

In order to have physically realistic estimates we take ${g}^{2}  
\cong {10}^{-1}$ in our calculations.  This leads to

\beq
\delta t \sim {8 \over h {\phi}_{0}} ~~~, ~~~ {t}_{d} \sim {m \over  
4{h}^{2}{{\phi}_{0}}^{2}}
\eeq
So resonance at each band of width $8hm{\phi}_{0}$
ends in a time $\sim{8 \over h{\phi}_{0}}$and in an
approximate time of ${m \over 4{h}^{2}{{\phi}_{0}}^{2}}$ decay  
effectively stops.  We notice that ${t}_{d}< {H}^{-1}$ for  
${\phi}_{0}{\  
\lower-1.2pt\vbox{\hbox{\rlap{$>$}\lower5pt\vbox{\hbox{$\sim$}}}}\  
}{({m \over h})}^{2}$.

In this analysis the effect of self-coupling of $\chi$
on the solutions of the Mathieu equation
was considered only to the first order, which is reasonable for the  
case of narrow-band resonance $h{\phi}_{0}{\  
\lower-1.2pt\vbox{\hbox{\rlap{$<$}\lower5pt\vbox{\hbox{$\sim$}}}}\ }  
m$.  The
key assumption in our treatment of the Mathieu equation in the  
presence of the nonlinear term was adiabaticity, i.e.
that we can use the instantaneous value of ${{m}^{2}}_{eff}$ which is  
also legitimate since
it changes over a time
$\delta t\sim {8 \over h{\phi}_{0}}$ which is greater than  
${m}^{-1}$, the period of oscillations of $\phi$, again because we  
are in the narrow-band resonance regime $h{\phi}_{0}{\  
\lower-1.2pt\vbox{\hbox{\rlap{$<$}\lower5pt\vbox{\hbox{$\sim$}}}}\ }  
m$.

Our results show major differences from the simple parametric  
amplification case.  The effect of large occupation number
for final state particles is not that dramatic here because there is  
a whole range of resonance bands
instead of a single one and the effect of the self-interaction of the  
produced particles drives modes out of the resonance bands much  
faster than the Hubble expansion.  Consequently the leading effect  
that influences the decay is the self-interaction of
the decay products, which stops it very early.

So far we have not considered thermalization of decay products and  
the Hubble expansion.  The temperature of the thermal bath after  
thermalization of $\chi$'s is calculated from
\footnote{In general the number density of particles is not the same  
before and after thermalization but the energy density is conserved.}

\beq
{{\pi}^{2} \over 30}({g}_{B}+{7 \over 8}{g}_{F}){T}^{4} =  
{\rho}_{\chi} = {n}_{\chi}{E}_{\chi} \sim  {{m}^{4} \over 16 {g}^{2}}
\eeq
where ${g}_{B}$ , ${g}_{F}$ are the number of bosonic and fermionic  
degrees of freedom, respectively.  We take the number of degrees of  
freedom to be the one in the minimal supersymmetric standatd model  
(${g}_{B}={g}_{F}=74$) which leads to a temperature $T\simeq {g}^{-{1  
\over 2}}{m \over 5}$.  For this temperature, however, the correction  
to mass-squared of $\chi$ is

\beq
{{m}^{2}}_{eff} \sim {g}^{2} {T}^{2} \sim g{{m}^{2} \over 25}
\eeq
which is much less than ${({m \over 2})}^{2}$ for ${g}^{2} \cong  
{10}^{-1}$.  This is just what we expected because ${{m}^{2}}_{eff}  
\sim {g}^{2}{{n}_{\chi} \over {E}_{\chi}}$ will be smaller after  
thermalization when the number density decreases and the mean energy  
of particles increases.  Therefore if thermalization is effective, it  
will lower ${{m}^{2}}_{eff}$ significantly which leads to further  
decay and thermalization.  This sequence of decay and thermalization  
stops when ${g}^{2} {T}^{2} \sim {{m}^{2} \over 4}$ that is at a  
temperature $T\sim m$ for ${g}^{2} \cong {10}^{-1}$ if the sequence  
is completed within a Hubble time.

Considering thermalization of decay products and the Hubble  
expansion, there are different possibilities depending on the  
relation among different time scales involved in the problem,  
${t}_{osc}\sim {m}^{-1}$, $\delta t\sim {8 \over h {\phi}_{0}}$,  
${t}_{d}\sim {m \over 4{h}^{2}{{\phi}_{0}}^{2}}$, ${t}_{th}\sim {32  
\pi \over \alpha m}$,
\footnote{${t}_{th}={{\Gamma}_{th}}^{-1}$ with ${\Gamma}_{th}\sim  
{\alpha}^{2} {{n}_{\chi} \over {m}^{2}}$ is for an out of equilibrium  
distribution of $\chi$'s.  It is easily seen that for a thermal bath  
with $T\sim {m \over 2 g}$ (which is the highest temperature that can  
be achieved) ${m}^{-1}{\  
\lower-1.2pt\vbox{\hbox{\rlap{$<$}\lower5pt\vbox{\hbox{$\sim$}}}}\  
}{t}_{th}$ also.}
and ${t}_{H}\sim\sqrt{{3 \over 8 \pi}}{1 \over m {\phi}_{0}}$.  The  
most important thing is that ${t}_{osc}$ is considerably smaller than  
all other time scales as long as we are in the narrow band regime  
${\phi}_{0}{\  
\lower-1.2pt\vbox{\hbox{\rlap{$<$}\lower5pt\vbox{\hbox{$\sim$}}}}\ }  
{m \over h}$.  Therefore changes in ${{m}^{2}}_{eff}$ caused by decay  
or thermalization (that can be considered as changes in the  
background) are adiabatic and our analysis is in principle valid,  
irrespective of the relation among ${t}_{d}$, ${t}_{th}$, ${t}_{H}$.   
Regarding these time scales there are different cases:

[1]- $\delta t{\  
\lower-1.2pt\vbox{\hbox{\rlap{$<$}\lower5pt\vbox{\hbox{$\sim$}}}}\ }  
{t}_{d}{\  
\lower-1.2pt\vbox{\hbox{\rlap{$<$}\lower5pt\vbox{\hbox{$\sim$}}}}\ }  
{t}_{th}{\  
\lower-1.2pt\vbox{\hbox{\rlap{$<$}\lower5pt\vbox{\hbox{$\sim$}}}}\ }  
{t}_{H}$.  This occurs when the inequalities ${\phi}_{0}{\  
\lower-1.2pt\vbox{\hbox{\rlap{$>$}\lower5pt\vbox{\hbox{$\sim$}}}}\ }  
5\times {10}^{-3}{m \over h}$ (to have ${t}_{d}{\  
\lower-1.2pt\vbox{\hbox{\rlap{$<$}\lower5pt\vbox{\hbox{$\sim$}}}}\ }  
{t}_{th}$), ${\phi}_{0}{\  
\lower-1.2pt\vbox{\hbox{\rlap{$<$}\lower5pt\vbox{\hbox{$\sim$}}}}\  
}3\times {10}^{-5}$ (to have ${t}_{th}{\  
\lower-1.2pt\vbox{\hbox{\rlap{$<$}\lower5pt\vbox{\hbox{$\sim$}}}}\ }  
{t}_{H}$) and ${({m \over h})}^{2}{\  
\lower-1.2pt\vbox{\hbox{\rlap{$<$}\lower5pt\vbox{\hbox{$\sim$}}}}\  
}{\phi}_{0}{\  
\lower-1.2pt\vbox{\hbox{\rlap{$<$}\lower5pt\vbox{\hbox{$\sim$}}}}\  
}{m \over h}$ (to have ${t}_{d}{\  
\lower-1.2pt\vbox{\hbox{\rlap{$<$}\lower5pt\vbox{\hbox{$\sim$}}}}\ }  
{t}_{H}$) are satisfied which necessarily means ${m \over h}{\  
\lower-1.2pt\vbox{\hbox{\rlap{$<$}\lower5pt\vbox{\hbox{$\sim$}}}}\ }  
{10}^{-2}$.  In this case the following sequence of events happens:  
decay of the inflaton to $\chi$'s, end of the decay, and  
thermalization of decay products to a temperature $T\sim m$, all in a  
time scale shorter than the Hubble time.  Hubble expansion then  
redshifts $T$ but for ${\phi}_{0}{\  
\lower-1.2pt\vbox{\hbox{\rlap{$>$}\lower5pt\vbox{\hbox{$\sim$}}}}\  
}{({m \over h})}^{2}$ (assuming ${t}_{d}{\  
\lower-1.2pt\vbox{\hbox{\rlap{$<$}\lower5pt\vbox{\hbox{$\sim$}}}}\  
}{t}_{th}{\  
\lower-1.2pt\vbox{\hbox{\rlap{$<$}\lower5pt\vbox{\hbox{$\sim$}}}}\  
}{t}_{H}$ all through the way) this sequence keeps $T\sim m$.  For  
${\phi}_{0}<{({m \over h})}^{2}$ decay is no longer effective to  
compensate for expansion and we must wait until a later time when  
${\phi}_{0}{\  
\lower-1.2pt\vbox{\hbox{\rlap{$<$}\lower5pt\vbox{\hbox{$\sim$}}}}\ }  
{h}^{2}$ and decay is completed as in the standard picture.  This  
second stage of decay dilutes the gravitinos that are produced  
earlier during the first stage.

[2]- $\delta t{\  
\lower-1.2pt\vbox{\hbox{\rlap{$<$}\lower5pt\vbox{\hbox{$\sim$}}}}\ }  
{t}_{th}{\  
\lower-1.2pt\vbox{\hbox{\rlap{$<$}\lower5pt\vbox{\hbox{$\sim$}}}}\ }  
{t}_{d}{\  
\lower-1.2pt\vbox{\hbox{\rlap{$<$}\lower5pt\vbox{\hbox{$\sim$}}}}\ }  
{t}_{H}$.  This occurs for ${\phi}_{0}{\  
\lower-1.2pt\vbox{\hbox{\rlap{$<$}\lower5pt\vbox{\hbox{$\sim$}}}}\  
}~min[5\times {10}^{-3}{m \over h},3\times {10}^{-5}]$ and ${({m  
\over h})}^{2}{\  
\lower-1.2pt\vbox{\hbox{\rlap{$<$}\lower5pt\vbox{\hbox{$\sim$}}}}\  
}{\phi}_{0}{\  
\lower-1.2pt\vbox{\hbox{\rlap{$<$}\lower5pt\vbox{\hbox{$\sim$}}}}\  
}{m \over h}$ which is possible only for ${m \over h}{\  
\lower-1.2pt\vbox{\hbox{\rlap{$<$}\lower5pt\vbox{\hbox{$\sim$}}}}\ }  
5\times {10}^{-3}$.  This case is similar to case [1], only because  
of the faster thermalization the temperature is higher and closer to  
the maximum possible $\sim m$.

[3]- $\delta t{\  
\lower-1.2pt\vbox{\hbox{\rlap{$<$}\lower5pt\vbox{\hbox{$\sim$}}}}\ }  
{t}_{d}{\  
\lower-1.2pt\vbox{\hbox{\rlap{$<$}\lower5pt\vbox{\hbox{$\sim$}}}}\ }  
{t}_{H}{\  
\lower-1.2pt\vbox{\hbox{\rlap{$<$}\lower5pt\vbox{\hbox{$\sim$}}}}\ }  
{t}_{th}$.  This occurs when ${\phi}_{0}{\  
\lower-1.2pt\vbox{\hbox{\rlap{$>$}\lower5pt\vbox{\hbox{$\sim$}}}}\  
}~Max[5\times {10}^{-3}{m \over h},3\times {10}^{-5}]$ and ${({m  
\over h})}^{2}{\  
\lower-1.2pt\vbox{\hbox{\rlap{$<$}\lower5pt\vbox{\hbox{$\sim$}}}}\  
}{\phi}_{0}{\  
\lower-1.2pt\vbox{\hbox{\rlap{$<$}\lower5pt\vbox{\hbox{$\sim$}}}}\  
}{m \over h}$ that is consistent only for ${m \over h}{\  
\lower-1.2pt\vbox{\hbox{\rlap{$>$}\lower5pt\vbox{\hbox{$\sim$}}}}\ }  
3\times {10}^{-5}$.  In this case decay stops without effective  
thermalization in a Hubble time, so the distribution of $\chi$'s is  
out of equilibrium.  Hubble expansion redshifts ${{m}^{2}}_{eff} $ as  
${R}^{-2}$ and further decay compensates for this change.  A thermal  
distribution of particles is not achieved, however, unless  
${t}_{th}{\  
\lower-1.2pt\vbox{\hbox{\rlap{$<$}\lower5pt\vbox{\hbox{$\sim$}}}}\ }  
{t}_{H}$
\footnote{In this case Hubble expansion moves particles from one band  
to another one.  This slightly lowers the time $\delta t$ spent in  
those bands because now there is an initial number of particles in  
each band.  This, however, is negligible since $\delta {n}_{\chi}$ is  
also redshifted as ${R}^{-3}$.}
.  For ${\phi}_{0}{\  
\lower-1.2pt\vbox{\hbox{\rlap{$<$}\lower5pt\vbox{\hbox{$\sim$}}}}\ }  
{({m \over h})}^{2}$ the situation is as in case [1].

[4]- ${t}_{th}{\  
\lower-1.2pt\vbox{\hbox{\rlap{$<$}\lower5pt\vbox{\hbox{$\sim$}}}}\ }  
\delta t{\  
\lower-1.2pt\vbox{\hbox{\rlap{$<$}\lower5pt\vbox{\hbox{$\sim$}}}}\ }  
{t}_{d}{\  
\lower-1.2pt\vbox{\hbox{\rlap{$<$}\lower5pt\vbox{\hbox{$\sim$}}}}\ }  
{t}_{H}$.  This occurs when ${\phi}_{0}{\  
\lower-1.2pt\vbox{\hbox{\rlap{$<$}\lower5pt\vbox{\hbox{$\sim$}}}}\  
}{10}^{-3}{m \over h}$ and ${({m \over h})}^{2}{\  
\lower-1.2pt\vbox{\hbox{\rlap{$<$}\lower5pt\vbox{\hbox{$\sim$}}}}\  
}{\phi}_{0}{\  
\lower-1.2pt\vbox{\hbox{\rlap{$<$}\lower5pt\vbox{\hbox{$\sim$}}}}\  
}{m \over h}$ which is possible only for ${m \over h}{\  
\lower-1.2pt\vbox{\hbox{\rlap{$>$}\lower5pt\vbox{\hbox{$\sim$}}}}\ }  
{10}^{-3}$.  In this case the decay products thermalize almost  
instantaneously and there is thermal equilibrium from the very  
beginning of the decay.  Actually both $\delta t$ and ${t}_{d}$ are  
much greater than $\sim {15 \over h{\phi}_{0}}$ and $\sim {m \over  
2{h}^{2}{{\phi}_{0}}^{2}}$ in this case because rapid thermalization  
of decay products in the thermal bath keeps the occupation number at  
each resonance band below one.  This means that temperature can be  
much lower than its maximum $\sim m$ and the situation is very  
similar to the one in the standard picture of reheating.

Cases [1] and [2] are the most dangerous regarding the problem of  
gravitino production.  In these cases thermal equilibrium can in  
principle be achieved from the very beginning and can last until the  
time ${H}^{-1} = \sqrt{{3 \over 8 \pi}}{m}^{-3}{h}^{2}$.  Even in  
these cases the gravitino overproduction is not that serious since at  
most we have a thermal bath with temperature $T \sim m$ for  a time  
$t \sim \sqrt{{3 \over 8 \pi}}{m}^{-3}{h}^{2}$, much shorter than $t  
\sim {m}^{-2}$ (assuming $m\ll {({m \over h})}^{2}$) which is the  
case for a radiation-dominated universe with temperature $T \sim m$.   
Case [3], on the other hand, is the most secure since the  
distribution of $\chi$'s is out of equilibrium, a distribution with  
less mean energy per particle and higher number density compared with  
a thermal distribution.  Depending on the parameters of the model  
$m,h$ and strength of the self coupling ${g}^{2}$ one or all of these  
cases can happen for ${({m \over h})}^{2}{\  
\lower-1.2pt\vbox{\hbox{\rlap{$<$}\lower5pt\vbox{\hbox{$\sim$}}}}\ }  
{\phi}_{0}{\  
\lower-1.2pt\vbox{\hbox{\rlap{$<$}\lower5pt\vbox{\hbox{$\sim$}}}}\ }  
{m \over h}$ but the temperature is at most of order $m$ as long as  
the inflaton is in this range, and is redshifted after that.  Also  
the thermal and non-thermal corrections to the mass-squared of $\chi$  
are always ${\  
\lower-1.2pt\vbox{\hbox{\rlap{$<$}\lower5pt\vbox{\hbox{$\sim$}}}}\ }  
{{m}^{2} \over 4}$.

To compare our results with that of parametric amplification without  
the effect of final state self-interaction let us consider the case
$h={10}^{-6}$,$m={10}^{-7}$, where the standard picture predicts
${T}_{R}\sim 0.1{m}^{{1 \over 2}}h={10}^{8}$ GeV.  In the simple  
parametric amplification case almost all the energy density of the  
inflaton is converted into radiation once ${\phi}_{0}{\  
\lower-1.2pt\vbox{\hbox{\rlap{$<$}\lower5pt\vbox{\hbox{$\sim$}}}}\  
}{m \over h}={10}^{-1}$ and leads to a very high reheat temperature  
${T}_{R}\sim{10}^{15}-{10}^{16}$ GeV which, from the point of view of  
gravitino overproduction, is a disaster.
According to our present analysis we are in the case [3] in the above  
all the way from ${\phi}_{0}={m \over h}={10}^{-1}$ to  
${\phi}_{0}={({m \over h})}^{2}={10}^{-2}$.  This means that  
thermalization is not effective for ${10}^{-2}{\  
\lower-1.2pt\vbox{\hbox{\rlap{$<$}\lower5pt\vbox{\hbox{$\sim$}}}}\ }  
{\phi}_{0}{\  
\lower-1.2pt\vbox{\hbox{\rlap{$<$}\lower5pt\vbox{\hbox{$\sim$}}}}\ }  
{10}^{-1}$ and occurs much later when decay is no longer effective,  
so the temperature during the first stage of decay is actually lower  
than $\sim {10}^{12}$ GeV. This results in a gravitino number density  
after the first epoch of decay which is too large by a factor of at  
most $10^{12}$.  The inflaton decay is then completed via a second  
stage when ${\phi}_{0}{\  
\lower-1.2pt\vbox{\hbox{\rlap{$<$}\lower5pt\vbox{\hbox{$\sim$}}}}\  
}{{h}^{2} \over 75}\simeq {10}^{-14}$.  By this time the temperature  
of the thermal bath and the momentum of the relativistic particles  
that might have been produced during the first stage are redshifted  
by a factor ${({{10}^{-28} \over {10}^{-4}})}^{{1 \over 3}} =  
{10}^{-{8}}$.  The second stage will determine the effective reheat  
temperature to be $\sim {10}^{8}$ GeV, and releases a large amount of  
entropy that dilutes the gravitinos produced during the first stage  
of decay  by a factor of ${({{10}^{12} \over {10}^{{8}}})}^{3} =  
{10}^{12}$ which is now sufficient as a dilution factor.

As we noted previously, since the effect which we consider is a  
kinematical cutoff due to the large self-induced plasma mass of the  
strongly self-interacting final state decay products, we expect to  
also have similar effects in the broad-band case.  Studies of the  
broad band resonance case \cite{7,10,14} indicate efficient  
production of the decay products only for masses up to about an order  
of magnitude larger than the inflaton mass,  so that when the  
broad-band resonance has built up a sufficient density of the decay  
products ( typically in a non-thermal "preheat" distribution) that  
their self-induced plasma mass exceeds this range, the decay will be  
suppressed. As above the decay will subsequently proceed as  
thermalization and Hubble expansion reduce the number densities and  
the induced self-mass for the decay products dips into the accessible  
range, resulting in a regular, quasi steady-state transfer of energy  
into the decay products.

Finally, we contrast the effects considered herein with those  
considered by Khlebnikov and Tkachev \cite {13}, who studied the  
semi-classical non-linear effects of the inflaton decay coupling in a  
massless $\lambda {\phi^{4}} $ model.  Note that the effects they  
study have a  different and independent origin from those considered  
here.  The effects which we study are {\it specifically} due to final  
state self-interactions of the decay products which are different  
from, and larger (gauge strength) than, the inflaton decay coupling.  
If these final-state self-couplings are present, then we saw above  
that their effects acted to regulate the parametric amplification of  
inflaton decay.

 Returning to the narrow-band case, as treated above, it is useful to  
ask if there are viable models where we are in the narrow-band regime  
from the beginning of oscillations, and for which the analysis  
presented here provides quantitative, as well as qualitative  
guidance.  We note that in a simple chaotic inflation model with  
potential $V(\phi )={1 \over 2}{m}^{2}{\phi}^{2}$ inflation ends when  
$V"(\phi )\simeq 24 \pi V(\phi)$ which happens for $\phi {\  
\lower-1.2pt\vbox{\hbox{\rlap{$<$}\lower5pt\vbox{\hbox{$\sim$}}}}\  
}{10^{-1}}$.  In the case of primordial supersymmetric inflation or  
new inflationary models the amplitude of  post-inflation oscillations  
around the global minimium is generically substantially smaller than  
the Planck scale.  Depending on the parameters then, some of these  
models may satisfy the inequalities necessary to be in the  
narrow-band resonance regime, or even for the nonlinear effects to be  
negligible. Interestingly enough, it seems that in some viable  
supersymmetric models this is indeed the case, and the analysis we  
have here undertaken is quantitatively valid.  We will return to  
these issues elsewhere \cite{31}.

In conclusion, we have seen that the self-interaction of final state  
bosons of moderate
strength, that arises very naturally in supersymmetric models,
has an important impact on the decay of observable sector inflatons,  
besides producing a rapid thermalization
rate.  As a step towards improved understanding
of the reheating process we have considered a simple schematic model  
representing generic features of supersymmetric theories, with such a  
final state self-coupling,
and have shown that in the case of narrow-band resonance the outcome  
is qualitatively different from that of simple
parametric amplification.  Here inflaton decay
occurs during two stages: one stage that consists of successive steps  
of
partial decay, thermalization and expansion at early times which ends  
relatively soon, and a second stage as
in the standard picture that completes the
decay.  In the first stage because of the quasi-adiabaticity  we were  
able to show that the temperature and the amplitude of quantum  
fluctuations of final state
particles are at most of the order of the mass of the inflaton  
(approximately),
and the temperature is several orders of magnitude below the naive
predictions of parametric amplification.  The second stage of
decay then determines the final reheat temperature and releases a  
substantial amount of entropy; is of particular importance in order  
to dilute the previously produced gravitinos in realistic  
supersymmetric models.

\noindent {\it Note Added} \\
After this paper was submitted  a preprint appeared from Prokopec and  
Roos [hep-ph/9610400] who do lattice simulations of inflaton decay.  
In the case where the decay product field has final state  
self-interactions of moderate strength their simulations exhibit  
regulation of the broad-band resonance decay, in agreement with the  
general analytic arguments we have presented above for the broad-band  
case.

\newpage

\noindent {\bf Acknowledgements} \\
 We are deeply indebted to both John Ellis and Andrei Linde for  
enlightening discussions, thoughtful criticism, and ongoing  
collaboration.
This work was supported in part by  the Natural Sciences and
Engineering Research Council of Canada.  In addition, RA wishes to  
thank the Iranian Ministry of Culture and Higher Education for  
continuing support.

\end{document}